\documentclass[
aps,%
12pt,%
final,%
notitlepage,%
oneside,%
onecolumn,%
nobibnotes,%
nofootinbib,
superscriptaddress,%
noshowpacs,%
centertags]%
{revtex4}
\usepackage[russian]{babel}
\usepackage{epsfig}
\usepackage{graphics}

\begin{document}



\title{PHASE AND INTENSITY DISTRIBUTIONS OF INDIVIDUAL PULSES OF
PSR~B0950$\pmb{+}$08}

\affiliation{Pushchino Radio Astronomy Observatory, Astro Space
Center, Lebedev Physical Institute, Pushchino, Moscow oblast',
Russia}

\author{\firstname{T.~V.}~\surname{Smirnova}}
\email[Электронный адрес: ]{tania@prao.ru}\noaffiliation

\received{March 15, 2006}
\revised{May 14, 2006}

\begin{abstract}
The distribution of the intensities of individual pulses of
PSR~B0950$+$08 as a function of the longitudes at which they
appear is analyzed. The flux density of the pulsar at 111~MHz
varies strongly from day to day (by up to a factor of 13) due to
the passage of the radiation through the interstellar plasma
(interstellar scintillation). The intensities of individual pulses
can exceed the amplitude of the mean pulse profile, obtained by
accumulating 770 pulses, by more than an order of magnitude. The
intensity distribution along the mean profile is very different
for weak and strong pulses. The differential distribution function
for the intensities is a power law with index $n = {-}1.1 \pm
0.06$ up to peak flux densities for individual pulses of the order
of 160~Jy.
\end{abstract}

\maketitle

\section{INTRODUCTION}

The mean profile obtained by summing several thousand individual pulses is
a stable characteristic of a pulsar at a given frequency. In spite of the
stability of the mean profile, however, the pulsar's radiation is very variable
on a wide range of time scales: nanoseconds for the giant pulses of the
Crab Pulsar~[1], tens and hundreds of microseconds for microstructure~[2, 3],
and several to tens of milliseconds for subpulses in individual pulses.
Variability of the amplitude from pulse to pulse and longer-term variability
associated with subpulse drift, nulling, and the passage of the radiation
through the interstellar plasma (scintillation) is also observed.

We will investigate variations of the amplitudes of subpulses of
the mean profile of PSR~B0950$+$08 at 111~MHz at various
longitudes (phases). The importance of this type of analysis is
that different mechanisms used to explain the coherent radio
emission of pulsars as being due to plasma microinstabilities or
non-linear processes have different electric-field statistics.
This parameter is determined by the internal statistics of the
radiation in the region of an individual source, effects due to
spatial modulation associated with the superposition of the
radiation from (possibly) many sources, and effects due to the
propagation of the radiation from the source to the observer. The
theory of the stochastic growth of plasma instabilities (SGT)~[4]
predicts a logarithmic normal distribution for variations of the
electric field. Non-linear three-wave processes acting in the
presence of high electric fields above some critical value $E_c$
lead to a power-law distribution with $P(E)\propto E^{-\alpha}$,
where $\alpha =4{-}6$. The theory of self-organized criticality
(SOC)~[5] predicts a power-law distribution with an index close to
$\alpha = 1$. Analysis of variations in the intensities of
individual pulses for the three pulsars PSR~B0833-45~[6],
B1641-45, and B0950$+$08~[7] showed that their variability
corresponds to log-normal field statistics; i.e., it is consistent
with the predictions of stochastic-growth theory. We will show
below that the variations in the amplitudes of the subpulses of
PSR~B0950$+$08 at 111~MHz are not consistent with these
statistics, and can be described well by a power law.

We chose PSR~B0950$+$08 for this analysis because it is one of the
most powerful pulsars at meter wavelengths, with a flux density at 102.5~MHz of
$S = 2$~Jy~[8]. This pulsar displays strong linear polarisation at 111~MHz,
$P_{l} = (70 {-} 80) \%$~[9], a weak interpulse located 152$^{\circ}$ from
the main pulse, microstructure with a characteristic time scale of about
150~$\mu$s~[3], and low-level extended radiation~[10, 11].

\section{OBSERVATIONS AND PRELIMINARY REDUCTION OF THE DATA}

Our data for PSR~B0950$+$08 were obtained at 111.2~MHz on the BSA
radio telescope of the Pushchino Radio Astronomy Observatory
(Astro Space Center of the Lebedev Physical Institute) in two
series of observations, during September 8--October 14, 2001 and
August 16--September 10, 2004. Only linear polarization was
received. A multi-channel receiver with 64 channels each with a
bandwidth of 20~kHz was used. The time for each observing session
was 3.3~min, corresponding to an accumulation of 770 individual
pulses. The time resolution in the first series of observations
was 0.4096~ms, and in the second series 1.28~ms; the receiver time
constant was 0.3 and 3~ms in the first and second series,
respectively. The dispersion broadening in each 20~kHz channel at
the observing frequency was 0.35~ms.

The individual pulses in all channels within a window of 150~ms
(first series) or 400~ms (1.6 times the pulsar period; second
series) were recorded on a computer disk, synchronous with the
pre-calculated topocentric pulse arrival times. The signals were
then dispersion-compensated and any channels subject to
substantial interference removed. Further, the signals in all
channels were added after a preliminary reduction to a single gain
via appropriate normalization, such that the dispersions of the
noise in all the channels were equal to the value averaged over
all the channels. The mean pulse profile for each observing
session was obtained by adding all the individual pulses. The mean
background determined from a noise interval outside the window of
the pulse radiation was subtracted from each pulse, and the rms
deviations of the noise level from the mean level in this region
were determined, which were then used to calculate the mean
$\sigma_{N}$ for the observing session for the individual pulses.

For each observing session, we calculated the peak amplitude $A_{\textrm{max}}$,
the signal-to-noise ratio S/N (the ratio of the amplitude of the mean
profile to $\sigma_{N}$, derived using a noise interval outside the mean
pulse), and the ``energy'' in the mean pulse, derived by summing the
intensities within the mean profile with values $I > 4 \sigma_{N}$. Our
analysis of the individual pulses consisted of determining the positions
(phases) and amplitudes of the subpulses within each pulse, which were then used
to construct the amplitude distribution functions at various longitudes of
the mean pulse.

\section{INFLUENCE OF POLARIZATION AND SCINTILLATION}

\begin{figure}[]
\includegraphics[angle=270, scale=0.6]{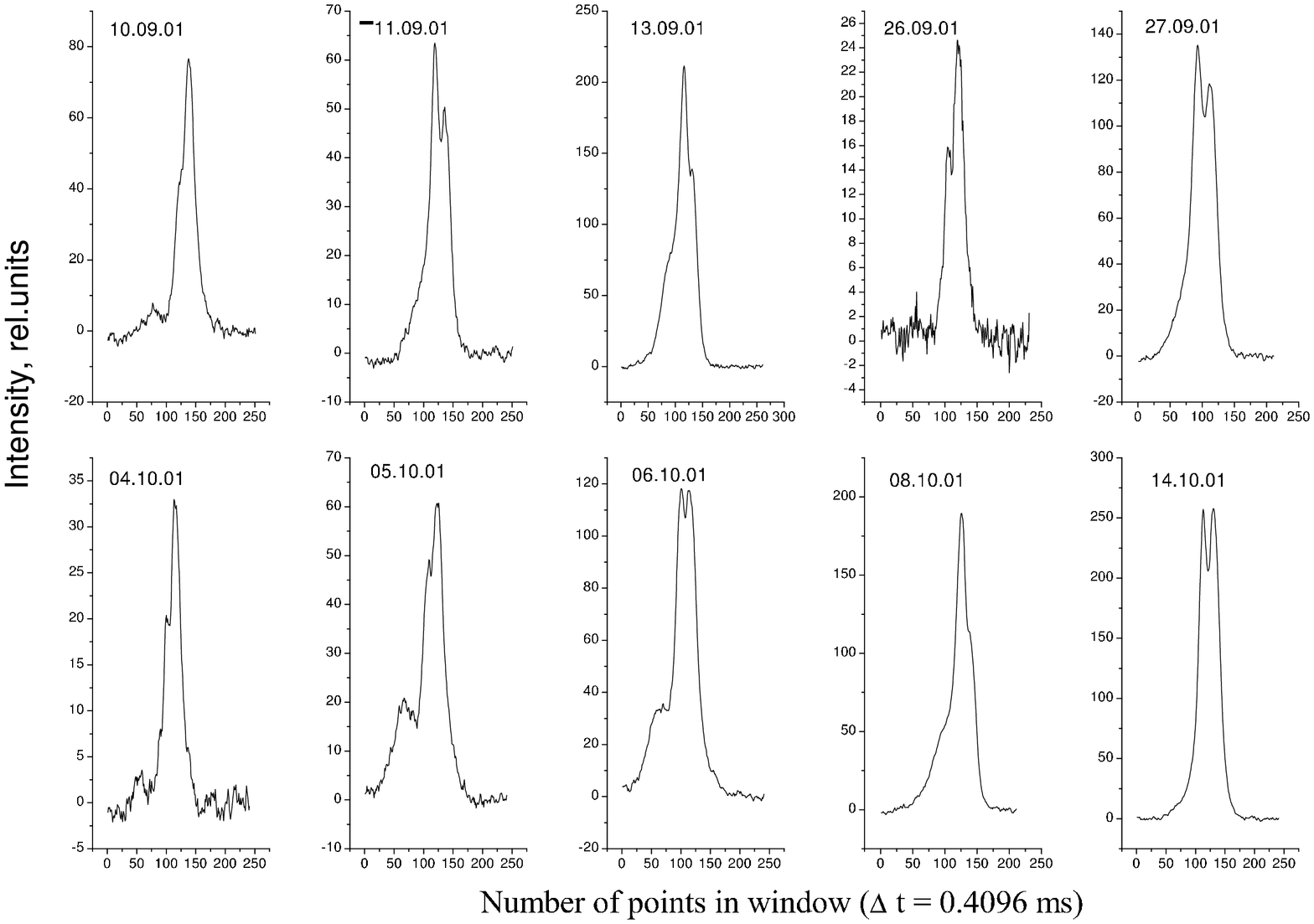
}
\setcaptionmargin{5mm}
\onelinecaptionstrue  
\caption{Mean pulse profiles for PSR~B0950$+$08 at $f=111.23$~MHz over 10
days in the first series of observations.\hfill}
\end{figure}

Figure~1 presents the mean profiles over 10 days in the first series of
observations. This figure clearly shows how strongly the pulse shape and
S/N varies from day to day. The profile varies from a well defined
three-component pulse (October 6, 2001) to a two-component pulse (October 14,
2001) in which the first component is virtually completely absent. This
behavior of the mean pulse reflects the strong influence of the polarization
of the received radiation. The analysis of profile variations in PSR~B0950$+$08
at frequencies 41--112~MHz carried out in~[9] showed that the rotation
measure for this pulsar is $\textrm{RM}= 4~{\textrm{rad}}/{\textrm{m}^{2}}$.
The period of the Faraday modulation at 111.2~MHz is 6~MHz, and the
corresponding rotation of the plane of polarization over the receiver bandwidth
($B = 1.28$~MHz) is 37$^{\circ}$. The contribution of the ionosphere to the
rotation measure at this frequency is no more than 10\%, and can be neglected.
The influence of the polarization of the received radiation amounts to
variations in the amplitudes of all three components from session to session.
The polarization profile at 151~MHz obtained by Lyne et al.~[12] shows
variations of the polarization angle by 160$^{\circ}$ along the mean profile,
which encompasses the unresolved (in those observations) first component and
the two others. Fitting polarization models to the observed frequency variations
of the mean-profile components yielded $100 \%$, $75 \%$, and $80 \%$ for the
degrees of linear polarization of the first, second and third components,
respectively~[9]. This means that the amplitude of the first component can
vary from zero to its maximum value, the second by a factor of two, and the
third by a factor of 2.4 from session to session. Accordingly, we must take
this effect into account when analyzing variations in the intensities of
subpulses at various longitudes of the mean profile.

PSR~B0950$+$08 is one of the most nearby pulsars, at a distance of $R =
262$~pc~[13] and with a dispersion measure of $\textrm{DM} =
2.97~{\textrm{~pc}}/{\textrm{cm}^{3}}$.  Accordingly, the pulsar radiation
should be strongly modulated by interstellar scintillation. A frequency
analysis of our data shows that the characteristic scale for the decorrelation
with frequency determined as the half-width of the autocorrelation function
(ACF) at the half-maximum level is $\Delta f_{d}$ = 200~kHz. The spectrum
does not vary over the observation time; i.e., the characteristic time scale
for the scintillations is longer than the observation time, $t_{d} > 3.3$~min.
The intensity in each spectral channel was determined by averaging the
intensities within a longitude interval chosen so that the signal in the mean
profile at these longitudes was more than half the maximum amplitude of the
mean pulse. We then averaged the spectra of 127 pulses, thus obtaining
spectra of the intensity variations every 32.1~s. The characteristic
decorrelation scale was determined from the mean ACF derived by averaging
the ACFs from spectra obtained over the entire observing session.

\begin{figure}[]
\includegraphics[scale=0.4]{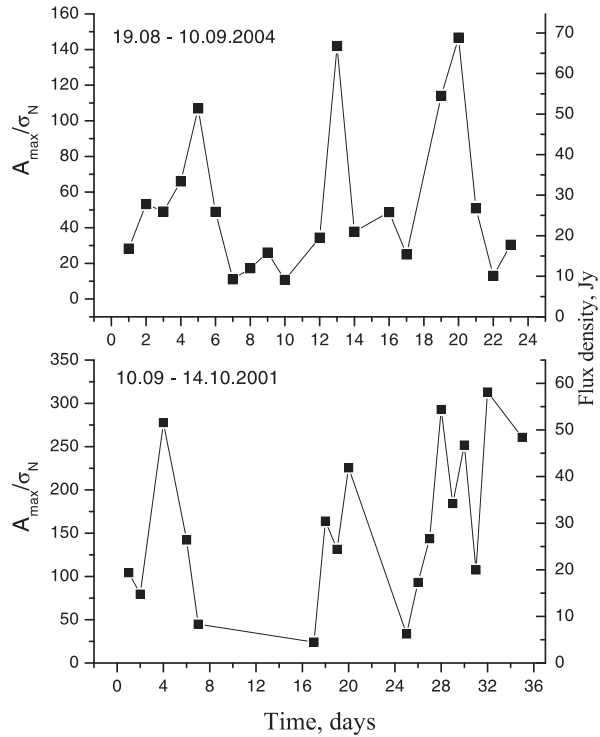}
\setcaptionmargin{5mm}
\onelinecaptionsfalse  
\caption{Time dependence of the amplitude of the mean profile in units of
$\sigma_N$ for the two series of observations. The peak flux-density scale
in Jy is indicated to the right.\hfill}
\end{figure}

Figure~2 presents the time dependence of the amplitude of the mean profile
$A_{\textrm{max}}$ in units of $\sigma_{N}$ for the two series of observations.
We can see the strong variability of this amplitude from day to day due to
scintillation. The maximum variations in the value of $A_{\textrm{max}}$ for
the mean profile reach a factor of 13. We used the following relation to
scale the peak pulse amplitudes for different days in flux-density units (Jy):
\begin{gather}
A_{\textrm{max}}(t) \mbox{~[Jy]} = \frac{A_{\textrm{max}}(t)}
{\langle A_{\textrm{max}}\rangle} \cdot S \cdot k_1,  \label{eq1}
\end{gather}
where $S = 2$~Jy is the flux density of the pulsar at our frequency (since
$f = 111.2$~MHz is close to the frequency of~[8], we neglected the frequency
dependence $S(f)$), $k_1 = 14.7$ is a coefficient relating the ratio of
the peak amplitude to the energy in the pulse averaged over the pulsar period,
and $\langle A_{\textrm{max}}\rangle$ is the mean of $ A_{\textrm{max}} $
over the entire series of observations in relative units. The energy in the
pulse is taken to be the sum of the intensities within the mean profile
with values $I > 4 \sigma_N$ multiplied by the time step between points.
For example, $A_{\textrm{max}}(t) \mbox{~[Jy]} = A_{\textrm{max}}(t) \cdot
0.187$ for the first and $A_{\textrm{max}}(t) \mbox{~[Jy]} =
A_{\textrm{max}}(t) \cdot 0.467$ for the second observation series. The
corresponding scale of the peak flux densities $S_p$ in Jy is shown to the
right in Fig.~2. The mean value of this quantity is $\langle S_p \rangle =
29.4$~Jy.

\begin{figure}[]
\includegraphics[angle=270,scale=0.5]{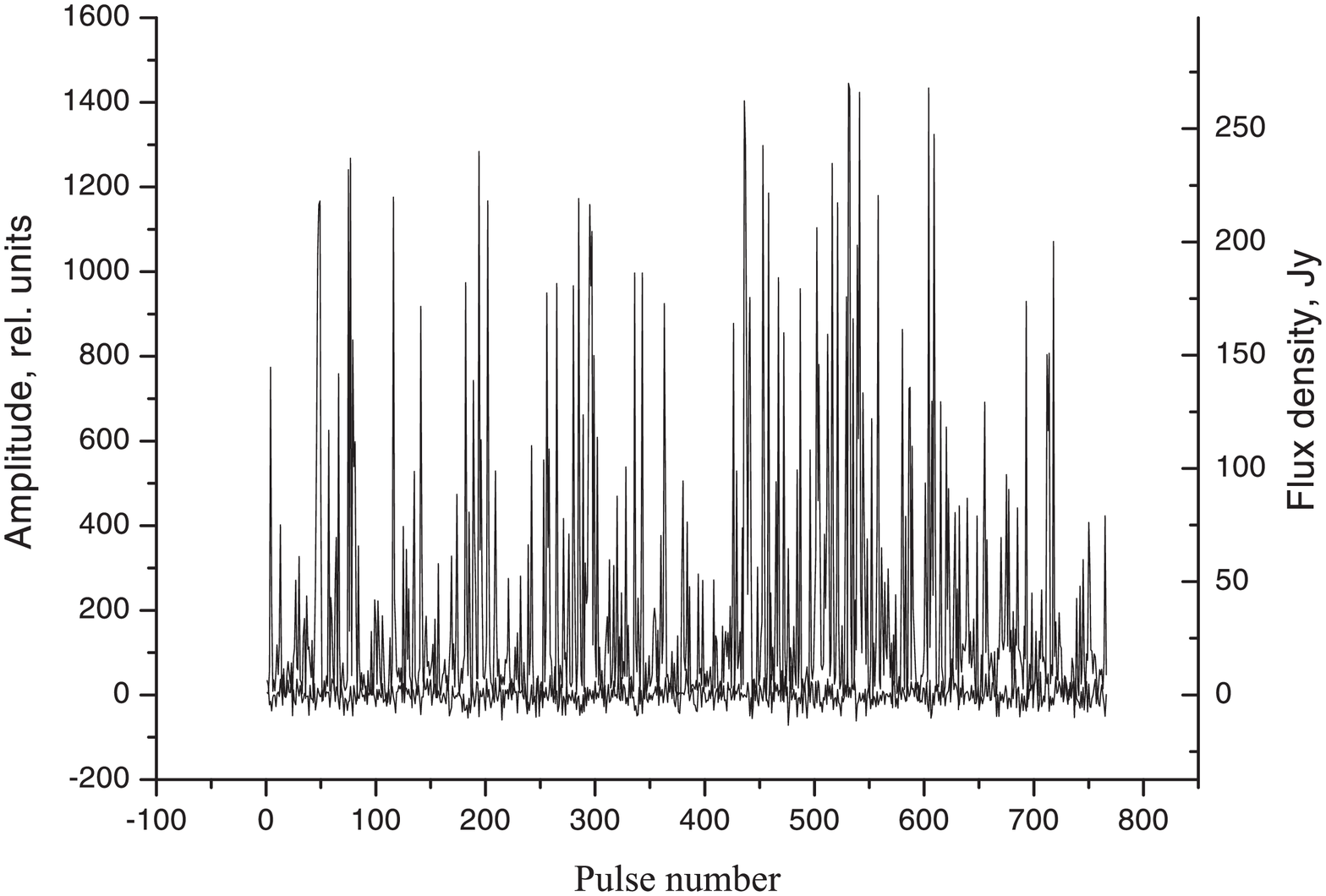}
\setcaptionmargin{5mm}
\onelinecaptionsfalse  
\caption{Variations in the amplitudes of individual pulses during the
observations of October 11, 2001. The variations of the mean noise amplitude
determined outside the emission window for these same pulses is also shown.
\hfill}
\end{figure}

\begin{figure}[]
\includegraphics[scale=0.6,angle=270]{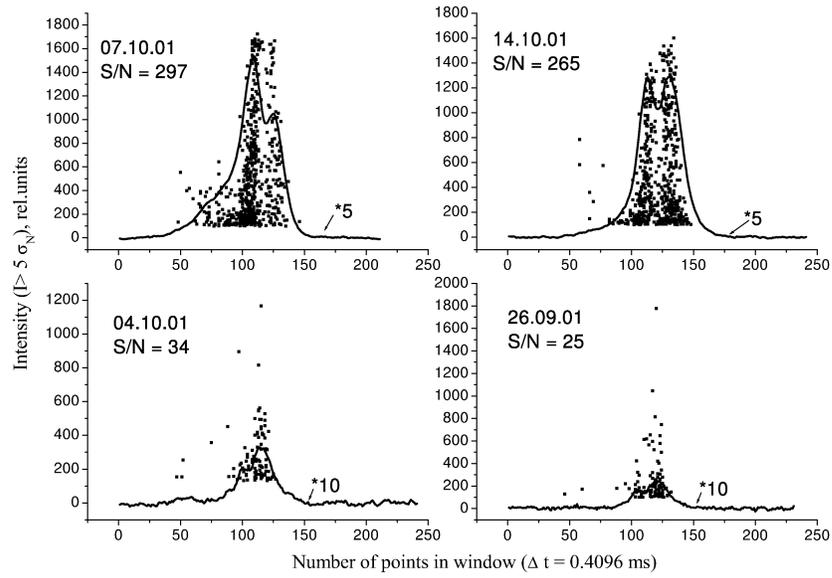}
\setcaptionmargin{5mm}
\onelinecaptionsfalse  
\caption{Distribution of the intensities of the individual pulses at various
longitudes of the mean pulse for several sessions in the first series of
observations. The mean pulse profile for a given session multiplied by
the indicated coefficient is also shown, as well as the signal-to-noise ratio
S/N for the corresponding mean profiles.\hfill}
\end{figure}

\begin{figure}[]
\includegraphics[scale=0.5]{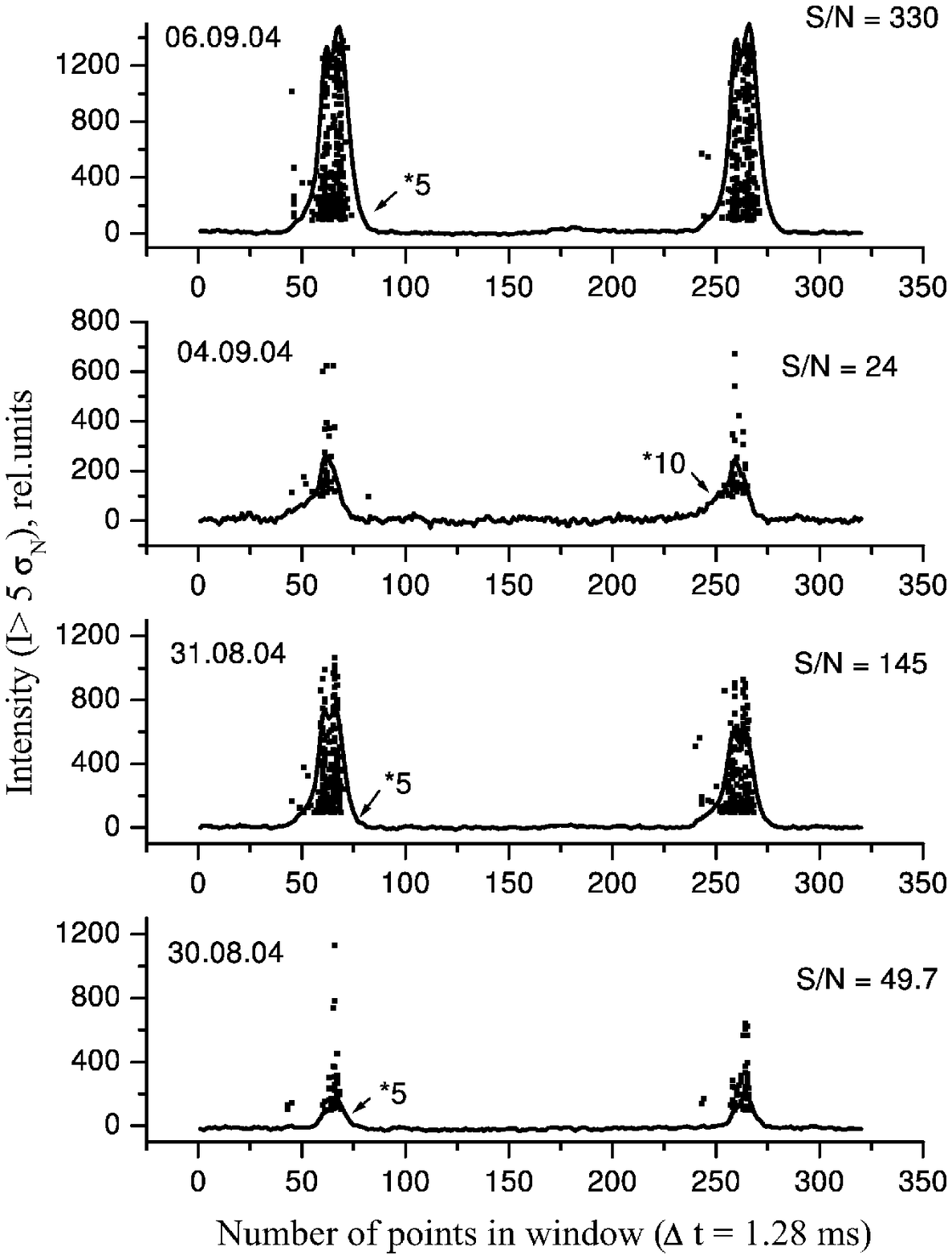}
\setcaptionmargin{5mm}
\onelinecaptionstrue  
\caption{Same as Fig.~4 for the second series of observations.
\hfill}
\end{figure}

\section{VARIATIONS OF THE INTENSITIES OF INDIVIDUAL PULSES}

Figure~3 shows the variations of the indiviual-pulse amplitudes
during the observing session on October 11, 2001. The amplitude
was taken at the longitude of the maximum of the mean profile. We
can clearly see the strong variability of the radiation, whose
intensity varies from zero to amplitudes with $\textrm{S/N}
\sim70$. The same figure presents the variations in the noise
amplitude outside the pulse-emission window.

Figures 4 and 5 show the longitude distributions of the peak amplitudes of
individual pulses over four sessions in the two series of observations. The
mean profiles multiplied by the indicated coefficients are also presented.
Here, we chose only pulses with amplitudes exceeding $5 \sigma_{N}$.
We can see three distinguished regions corresponding to the
positions of the maxima of the three components of the mean profile, where
subpulses appear most often. The maximum pulse amplitudes for sessions with
low S/N (here, $\textrm{S/N} = A_{\textrm{max}}/\sigma_N$ for the mean
profile) exceed the value $A_{\textrm{max}}$ for the mean profile by more than
an order of magnitude. For sessions with high S/N (${\sim}$300), the maximum
pulse amplitudes are close to $5 A_{\textrm{max}}$, while their absolute values
are approximately equal to the amplitudes of the strongest pulses for weak
records. This reflects the fact that the dynamic range of our analog--digital
converter is not sufficient to accurately register the amplitudes of the
strongest pulses, so that the amplitudes are cut off when the energy in the
pulse grows substantially due to scintillation. Therefore, the real intensities
of the strongest pulses are reflected by recordings with low mean-profile
amplitudes. We took this effect into account in our subsequent analysis of
the data. Note that the peak amplitude of the strongest pulse $S_p$ for
September 26, 2001 exceeds the amplitude of the mean profile for that session
by a factor of 60, which corresponds to $S_p \geq 270$~Jy.

\section{DISTRIBUTION FUNCTION}

The distribution of the number of pulses exceeding a given intensity threshold
(for two observing sessions) is presented in the upper part of Fig.~6 on a
double-log scale. The intensities of the individual pulses are normalized
using $\sigma_N$. We used only pulses with amplitudes exceeding $5 \sigma_N$
to construct this distribution. The mean profiles for the corresponding
sessions are also shown. We obtained separate distributions for pulses
appearing at the longitudes of the maxima of the first, second, and third
components on September 13, 2001 and of the second and third components on
October 14, 2001. We can see that the distribution becomes appreciably steeper
at high values of $I/\sigma_N$, which refects the cut-off of the amplitudes of
strong pulses due to the instrumental effect mentioned above. The vertical
lines in the figure correspond to the cut-off boundaries, which depend on
the S/N of the mean profile for a given session.

\begin{figure}[]
\includegraphics[scale=0.5]{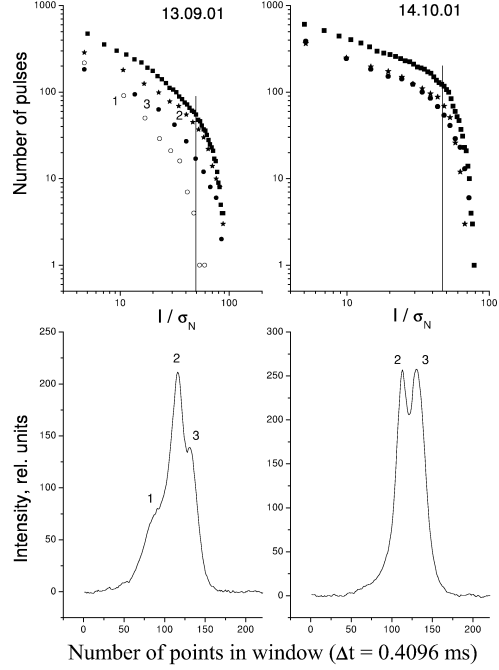}
\setcaptionmargin{5mm}
\onelinecaptionsfalse  
\caption{Number of pulses exceeding a specified intensity
threshold as a function of the intensity $I$ in units of
$\sigma_N$ (upper plots) together with the corresponding mean
profiles (lower plots). Distributions were constructed separately
for pulses appearing at the longitudes of the maxima of the first
component (hollow circles), second component (asterisks), and
third component (filled circles). The squares show the overall
distribution of the pulses independent of the longitude at which
they appear.\hfill}
\end{figure}

\begin{figure}[]
\includegraphics[scale=0.4,angle=270]{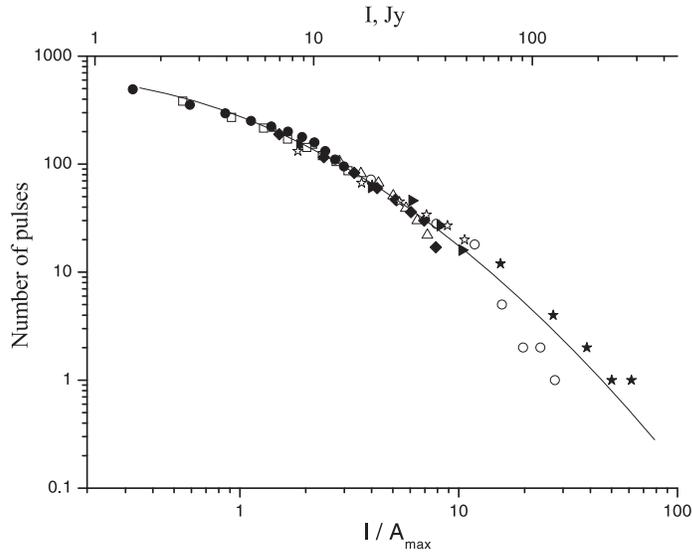}
\setcaptionmargin{5mm}
\onelinecaptionsfalse  
\caption{Integral distribution function constructed for eight days of
observations. The data for different days are shown by different symbols.
A parabola fit to all the points using the least-squares method is also
shown.\hfill}
\end{figure}

\begin{figure}[]
\includegraphics[scale=0.5,angle=270]{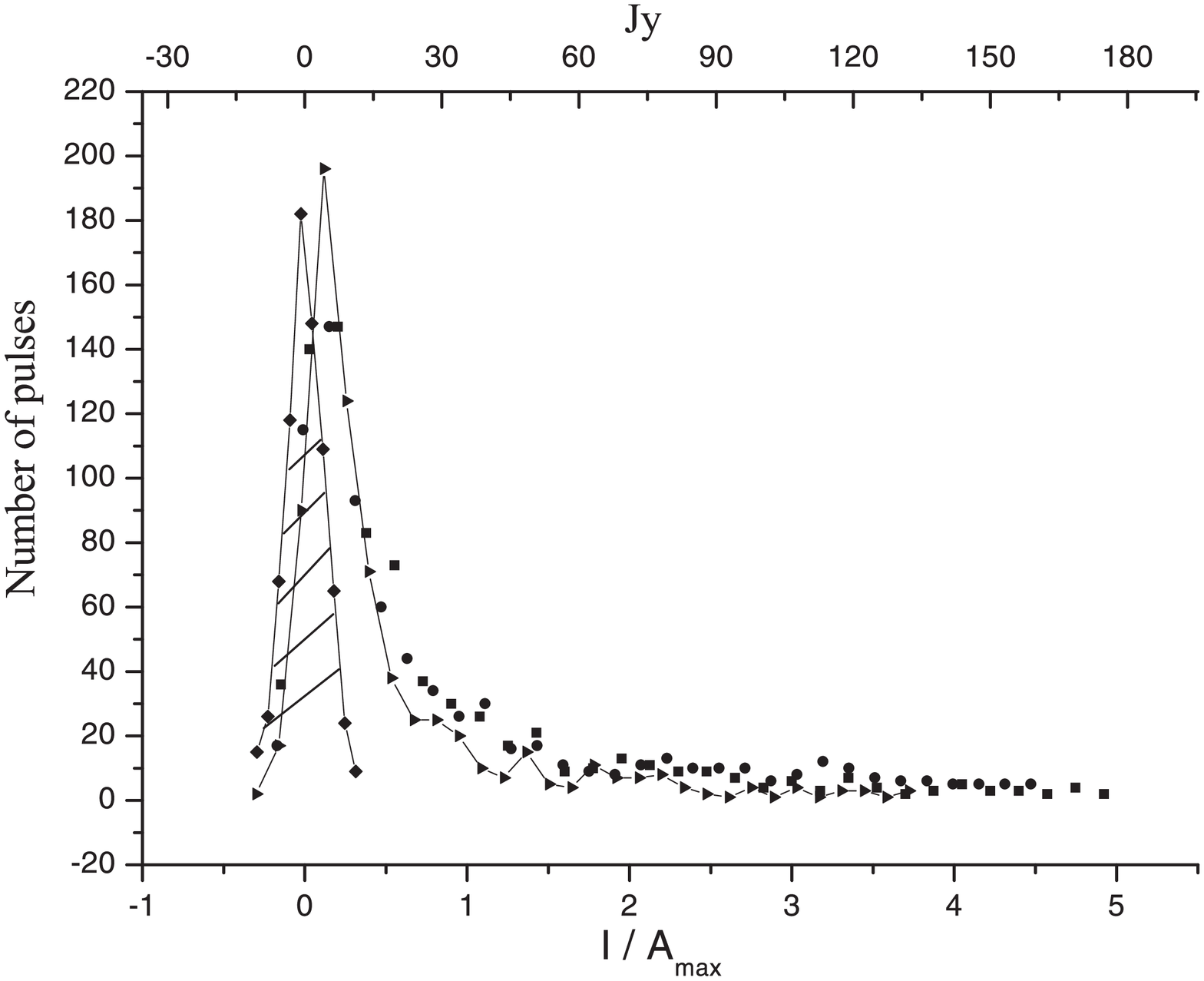}
\setcaptionmargin{5mm}
\onelinecaptionsfalse  
\caption{Differential distribution function constructed for the noise (shaded)
and the data for three days of observations with high signal-to-noise ratio
S/N in the mean profile. The intensity was taken at the longitude of the
maximum for the mean profile.\hfill}
\end{figure}

At the longitude of the maximum of the first component, the distribution
becomes steeper, only the strongest pulses reach the cut-off boundary, and
the distribution correctly reflects the original distribution. At the
longitudes of the second and third components, the distributions are similar
out to the cut-off boundaries, differing only in a coordinate shift
corresponding to the ratio of the amplitudes of these components in the mean
profile. On October 14, 2001, the component amplitudes are approximately the
same, and the distributions coincide. Since the strong linear polarization of
the components leads to substantial variations of their amplitudes in the mean
profile, we used the pulse intensities at the longitude of the mean-profile
maximum to construct the integral distribution functions using the data for
different days of observations. To exclude the effect of scintillation, we
normalized the pulse intensities to the amplitude of the mean profile,
$A_{\textrm{max}}$.

Figure~7 shows the resulting distributions constructed for eight days of
observations, with pulses with amplitudes exceeding the cut-off boundaries
excluded. We can see that the data for different days are in very good
agreement, testifying that we have correctly removed the indicated instrumental
effect and the influence of polarization and scintillation. The distributions
presented on a double-log scale are well described by a parabola, and we
obtained a parabolic least-squares fit to all the points in the distribution,
also shown in Fig.~7: $\log  N   = 2.44 - 0.77\log (I/A_{\textrm{max}}) -
0.43\log(I/A_{\textrm{max}})^2$.

Figure~8 presents differential distribution functions constructed using the
noise (left) and pulse data for three days of observations with high S/N.
The intensities of individual pulses were taken at the longitude of the maximum
of the mean profile. The noise distribution is Gaussian, with a zero mean and
$\sigma$ = 0.07 in units of $I/A_{\textrm{max}}$. When constructing the
pulse distribution, the only constraint we applied was excluding high
intensities that exceeded the cut-off boundary. The maximum of this
distribution is shifted from zero by $x = 0.117$. This indicates that there
are an appreciable fraction of pulses with small intensities in the radiation
of PSR~B0950$+$08.

\begin{figure}[]
\includegraphics[scale=0.4,angle=270]{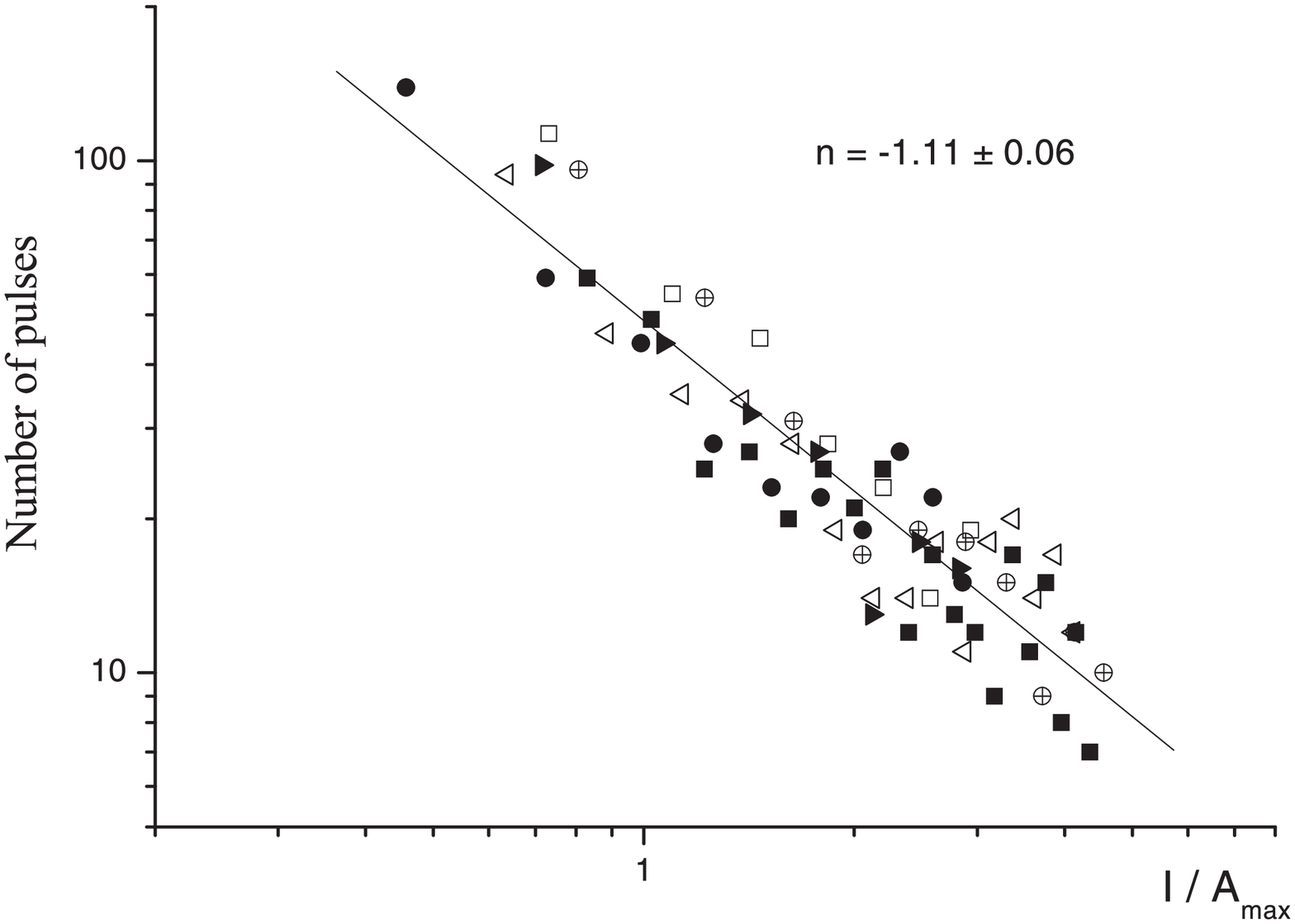}
\setcaptionmargin{5mm}
\onelinecaptionsfalse  
\caption{Differential distribution function constructed using data for six
days of observations shown on a double-log scale. The intensities were taken
at the longitude of the mean-profile maximum. The line is the result of a
least-squares fit.\hfill}
\end{figure}

\begin{figure}[]
\includegraphics[scale=0.5]{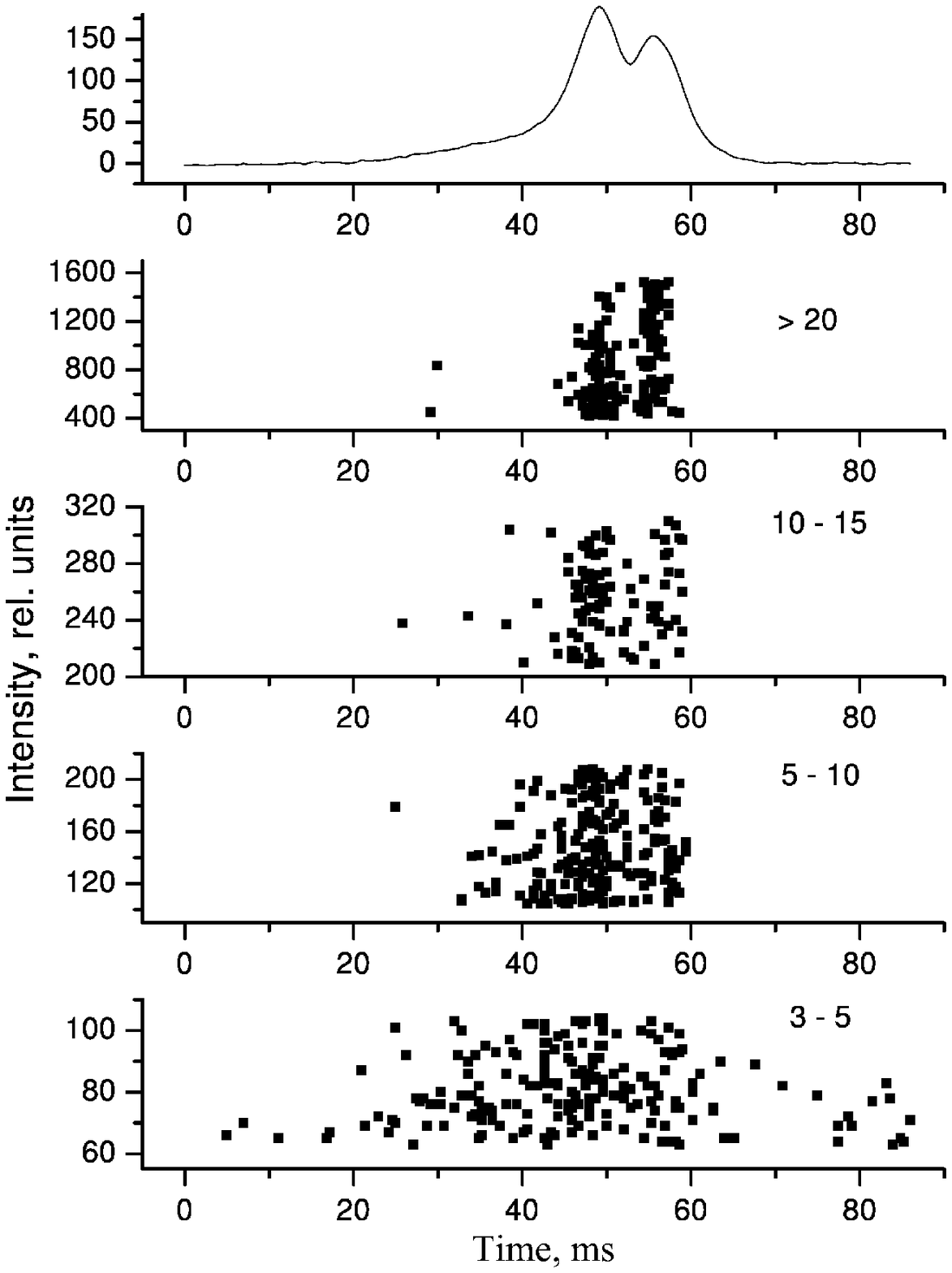}
\setcaptionmargin{5mm}
\onelinecaptionsfalse  
\caption{Longitude distribution of the individual pulses with the indicated
range of intensities in units of $\sigma_N$. The mean profile for October 11,
2001 is shown above.\hfill}
\end{figure}

The differential distribution function for pulses with $I > 5\sigma_N$ for
six days of observations with high S/N is presented in Fig.~9 on a double-log
scale. Here, we likewise used only data to the cut-off boundary. A linear
least-squares fit to all the points has the slope $n = {-}1.11 \pm 0.06$ to
intensities $I/A_{\textrm{max}} \approx$ 4.5 or 160~Jy. The distributions
for days with relatively low signal-to-noise ratios $\textrm{S/N} \approx
30{-}120$, i.e., with appreciably more distant cut-off boundaries, are the
steepest: $n = {-}1.5 \pm 0.2$ for $I > 160$~Jy. Although the statistics
for pulses with high intensities are relatively poor, it appears that these
results confirm a steepening of the distribution function for $I> 160$~Jy.

The power-law distribution function $P(I) \sim I^{-n}$ ($n \approx
1$) that we have obtained for the ordinary pulsed radiation of
PSR~0950+08 agrees with the predictions of self-organized
criticality theory~[5]. This theory is based on the idea that
there is a self-consistent interaction between waves, the flow of
moving particles, and the surrounding plasma, which is near
marginal stability. This theory predicts a power-law distribution
$P(E^2) \sim E^{-\beta}$ over a wide range of energies
(intensities), with $\beta \approx 1$ and varying from 0.5 to 2
for various systems~[5], and the power-law distribution $P(E) \sim
E^{-\alpha}$ for the electric field $E = \sqrt I $. Here, $\alpha
= 2\beta - 1$~[14], and accordingly $\alpha \approx 3.4$ and
$\beta \approx 2.2$ in our case. As was shown in~[14], giant
pulses and giant micropulses in a number of pulsars display
power-law distribution functions with $n = 4.4 {-} 6.5$,
appreciably higher than our value $n \approx$~1. This provides
evidence that giant pulses and micropulses are generated by a
different process than that giving rise to the normal radiation
and/or are formed in a very different region.

The statistics of the 430~MHz pulsed radiation of PSR~0950+08 at
various phases of the mean profile was studied by Cairns et
al.~[7], who showed that the distribution function for the
logarithm of the electric field, $P(\log E)$, varies significantly
with phase, with the distribution at the longitude of the
mean-profile maximum being approximately flat in the range $2.1
\leq \log E  \leq 2.7$ and falling off in accordance with a power
law at $\log E > 2.7$~[7, Fig.~22]. This range of $E$ corresponds
to their data for $5 \sigma_N \leq I \leq 75 \sigma_N$. Cairns et
al.~[7] fit a combination of a Gaussian intensity distribution and
a non-linear log-normal distribution to the distribution they
obtained. The statistics for the fit are relatively poor, but it
provides a good qualitative description of the data. Since $P(\log
E) = 2IP(I)$, the distribution we have obtained, $P(I) \sim
I^{-1.1}$, corresponds to a flat $P(\log E)$~[7]. We also observe
a steepening of $P(I)$ at high intensities. It appears that the
distribution of the intensity at the longitude of the mean-profile
maximum is described by a piecewise-power-law function.

\begin{figure}[]
\includegraphics[scale=0.7,angle=270]{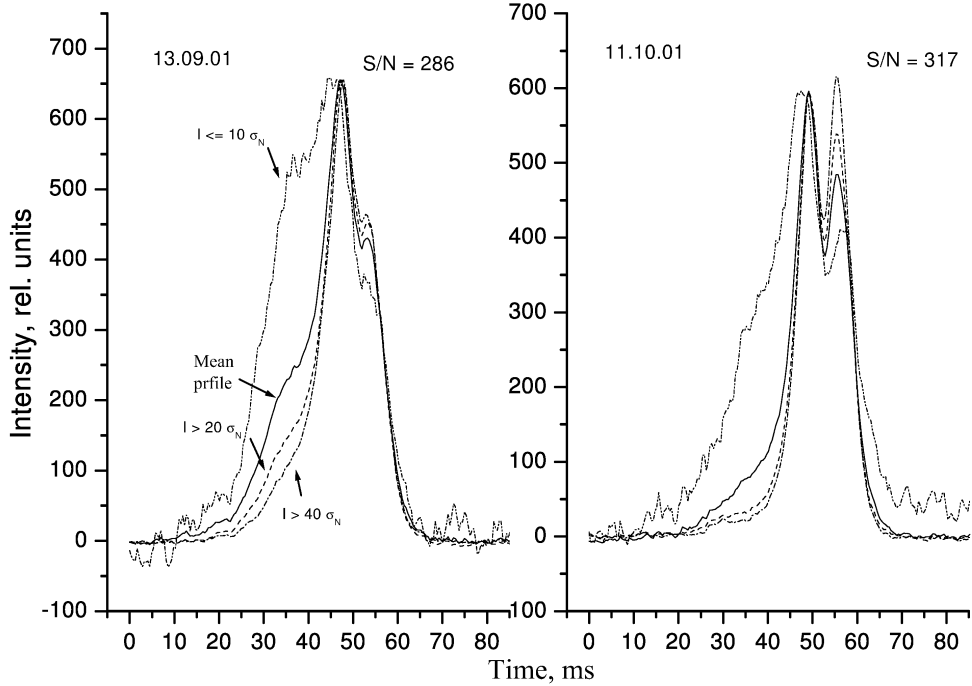}
\setcaptionmargin{5mm}
\onelinecaptionsfalse  
\caption{Mean profiles for two days of observations obtained by summing all
pulses (solid curve) and pulses with intensities  $3\sigma_N < I <  10\sigma_N$
(dotted curve), $I > 20 \sigma_N$ (dashed curve), and $I > 40 \sigma_N$
(dot--dashed curve). The amplitudes of all the profiles (in relative units)
have been normalized to a single value.\hfill}
\end{figure}

\section{LONGITUDE DISTRIBUTIONS FOR WEAK AND STRONG PULSES}

In order to investigate the longitude distributions of pulses with
different intensities, we divided the pulses into groups having
various intensity ranges: $(3-5)\sigma_N$, $(5-10)\sigma_N$,
$(10-15)\sigma_N$, and more than 20$\sigma_N$. As we can see in
Fig.~10, the weaker the pulses, the broader the range of
longitudes in which they appear. The stronger the pulses, the
narrower the longitude region in which they appear and the more
they are concentrated toward the longitudes of the maximum
intensities of the second and third components of the mean
profile. An analysis for many sessions indicates that this is a
general property of this pulsar that does not depend on the shape
of the mean profile.

Figure~11 shows mean profiles obtained by summing pulses with
intensities in the ranges indicated in the figure. We can see the
dependence of the longitude distribution of the intensities on the
intensity of the pulses used in the sum. In the case of the
profile obtained by summing weak pulses, the amplitude of the
first component relative to the second component grows
appreciably, while the amplitude of the third component is
decreased. The stronger the pulses that are summed, the lower the
contribution of the first component and the greater the amplitude
of the third component. This behavior is characteristic of all
sessions, and does not depend on the shape of the mean profile
obtained by summing all pulses. The width of the profile at the
half-maximum level, 0.5$A_{\textrm{max}}$, obtained by summing
weak pulses is approximately twice the width of the profiles
obtained by summing strong pulses, although the widths at the
level 0.1$A_{\textrm{max}}$ are approximately the same. Analysis
of the pulse intensities for the second series of observations,
when we recorded the pulsar emission in a 400~ms window (1.6
$P_1$), showed an absence of pulses with amplitudes exceeding
5$\sigma_N$ outside the range of longitudes for the main pulse.

\section{CONCLUSION}

Our analysis of variations in the pulse intensities for PSR~B0950$+$08
has shown the presence of strong variations in the amplitude of the mean
profile (by up to a factor of 13) due to diffractive scintillation, whose
time scale exceeds the time for individual observations ($T > 4$~min).
The intensities of individual pulses can exceed the amplitude of the mean
profile on individual days by more than an order of magnitude ($I > 340$~Jy).
Does PSR~B0950$+$08 display giant pulses? Although the strongest recorded
pulse exceeded the amplitude of the mean profile on that day by a factor of
60, it exceed the mean amplitude for the entire series of observations by
only a factor of 9.3. The strongest pulses appear within a narrow range of
longitudes near the longitude of the third component, and display a power-law
distribution. All of these properties are consistent with the characteristics
of giant pulses, although the relative amplitudes of the strongest pulses
are appreciably lower than those for the giant pulses of the Crab Pulsar.
The question of which pulses should be considered giant pulses remains open,
since there is no precise definition for this phenomenon. We also note that
searches for giant pulses from weak, nearby pulsars must take into account
the influence on the observed flux densities of effects associated with the
passage of the radiation through the interstellar plasma. If a pulsar is very
weak and its mean profile can be distinguished only after accumulating a large
number of pulses, individual pulses may be visible at specific times due to
the enhancement of the signal amplitude due to scintillation. These pulses
could be erroneously identified as giant pulses when compared with the low
amplitude for the mean pulse over a long time interval.

We have investigated the longitude distributions of pulses with
differing intensities. In the case of weak pulses, the radiation
at the longitude of the first component grows appreciably, while
strong pulses are primarily realized at the longitudes of the
second and third components of the mean profile. The radiation of
both weak and strong pulses probably comes from a single level in
the pulsar magnetosphere, since the total widths of their
radiation cones, indicated by the widths of their mean profiles at
a low level of the profile, are approximately the same. The
differential distribution function is well fit by a power law with
index $n = {-}1.1 \pm 0.06$, in agreement with the predictions of
SOC theory. There is some evidence that the distribution function
becomes steeper for pulse intensities exceeding 160~Jy.

\section{ACKNOWLEDGEMENTS}

This work was supported by the Russian Foundation for Basic Research
(project codes 03-02-16522, 03-02-16509) and the National Science Foundation
(grant AST0098685).

\end{document}